\documentclass[aps, prl, reprint, twocolumn, superscriptaddress]{revtex4-1}
\usepackage{multirow}
\usepackage{graphicx}
\usepackage{longtable}
\usepackage[utf8]{inputenc}
\usepackage[T1]{fontenc}
\usepackage{epstopdf}
\usepackage{textcomp}
\usepackage[cmyk,dvipsnames]{xcolor} 
\usepackage{amsbsy}
\usepackage{amsmath}
\usepackage{amssymb}
\usepackage{amsfonts}
\usepackage{dsfont}
\usepackage{mathtools}
\usepackage{braket}
\usepackage{array}
\usepackage{gensymb}
\usepackage{placeins}
\usepackage{dashrule}
\usepackage{xcolor}
\usepackage{booktabs}
\usepackage{hyperref}
\usepackage{float}
\usepackage{braket}
\usepackage{ulem}

\definecolor{ao(english)}{rgb}{0.0, 0.5, 0.0}

\makeatletter
  \def\my@tag@font{\normalsize}
  \def\maketag@@@#1{\hbox{\m@th\normalfont\my@tag@font#1}}
  \let\amsmath@eqref\eqref
  \renewcommand\eqref[1]{{\let\my@tag@font\relax\amsmath@eqref{#1}}}
\makeatother

\allowdisplaybreaks

\begin{document}
\title{Coupled quasimonopoles in chiral magnets}

\newcommand{\fz}{Peter Gr\"unberg Institut and Institute for Advanced Simulation, Forschungszentrum J\"ulich and JARA, 52425 J\"ulich, Germany}
\newcommand{\iceland}{Science Institute and Faculty of Physical Sciences, University of Iceland, 107 Reykjav\'{i}k, Iceland}
\newcommand{\rwth}{Department of Physics, RWTH Aachen University, 52056 Aachen, Germany}
\newcommand{\Sweden}{Department of Physics, KTH-Royal Institute of Technology, SE-10691 Stockholm,  Sweden}

\author{Gideon P. M\"uller}
 	\homepage[]{g.mueller@fz-juelich.de}
 	\affiliation{\fz}
	\affiliation{\iceland}
	\affiliation{\rwth}
\author{Filipp N. Rybakov}
	\affiliation{\Sweden}
\author{Hannes J\'onsson}
	\affiliation{\iceland}
\author{Stefan Bl\"ugel}
	\affiliation{\fz}
\author{Nikolai S. Kiselev}
	\email[]{n.kiselev@fz-juelich.de}
	\affiliation{\fz}

\date{\today}

\begin{abstract}
Magnetic singularities, also known as magnetic monopoles or Bloch points, represent intriguing phenomena in nanomagnetism.
We show that a pair of coupled Bloch points may appear as a localized, stable state in cubic chiral magnets.
Detailed analysis is presented of the stability of such objects in the interior of crystals and in geometrically confined systems.
\end{abstract}

\maketitle
%
%
Point singularities in magnets~\cite{Feldtkeller_65,Doring_68} are topologically non-trivial objects known as Bloch points (BPs)~\cite{Malozemoff_79} or hedgehogs and recently have become commonly referred to as quasi-monopoles or monopoles~\cite{Milde,Kanazawa_16}.
Within the classical continuum micromagnetic approach, such a singularity is a point where the  magnetization vector $\mathbf{M}(\mathbf{r})$ suffers a discontinuity, but there is, nevertheless, no divergence in the energy~\cite{Doring_68}.
Contrary to the hypothetical Dirac monopole, the magnetic flux through the surface surrounding a magnetic singularity is equal to zero.
Nevertheless, noticeable similarity between these objects does exist.
For instance, similar to the Dirac string stretching from a magnetic monopole, a magnetic singularity can be considered as an origin of a skyrmion string~\cite{Milde} representing strings-vortices in the magnetization vector field.
Typically, skyrmion strings are observed as singularity-free magnetic textures stretching throughout the entire sample between free edges.
Of particular interest are systems where one or both ends of the string behave differently and instead of being stretched to a free surface of the sample are coupled to point singularities.
The scenario in which one end of a skyrmion string ends at a surface of the crystal but the other ends with a point singularity within the sample has been discussed earlier for isotropic chiral magnets~\cite{Schutte}. 
This kind of texture may become stable and underlies a new type of particle-like state –- the chiral bobber~\cite{Rybakov_15}. It has recently been observed by means of off-axis electron holography~\cite{Zheng_18}.
Such objects composed of a point singularity coupled to a continuous vector field can be classified as so-called hybrid solitons~\cite{Rybakov_15}.
Another possible scenario corresponds to the case when both ends of the string represent point singularities with opposite topological charge.
The stability of such a magnetic texture has been assumed to require the singularities to be located in the vicinity of opposite surfaces of the sample~\cite{Leonov_18}.
The mechanism of stabilization of such a configuration is similar to that of a coupled pair of chiral bobbers situated near opposite surfaces of the film~\cite{Rybakov_16}.

A more sophisticated mechanism of stabilization of such a configuration has been discussed in Ref.~\cite{Liu_Binding_2018}, where the particular case of a thin disk of a chiral magnet is discussed.
The disk is covered by additional magnetic layers characterized by strong perpendicular magnetic anisotropy and coupled to a chiral magnet via exchange interaction. 
The introduction of such artificial interfaces serve to pin the spins on the top and bottom surfaces of the chiral magnet to be pointed perpendicular to the interface. 
Such a method of artificial confinement is analogous to the stabilization of exotic textures in liquid crystals squeezed between glass plates where the surfaces are treated to align the molecules along the normal~\cite{Smalyukh_2010}.

We present in this Letter results showing that the stability of such a coupled pair of quasi-monopoles does not in general require geometric confinement but can be achieved even within the interior of bulk chiral magnets
favoring skyrmions~\cite{Bogdanov_89}, such as MnSi~\cite{Yu_15}, FeGe~\cite{Yu_11}, Fe$_{1-x}$Co$_x$Si~\cite{Yu_10} and some other Si- and Ge-based metallic alloys and insulating magnets such as Cu$_{2}$OSeO$_{3}$~\cite{Seki_2012} and alloys of $\beta$-Mn-type Co-Zn-Mn~\cite{Tokunaga_15}.

\begin{figure}[ht]
    \includegraphics[width=0.9\linewidth]{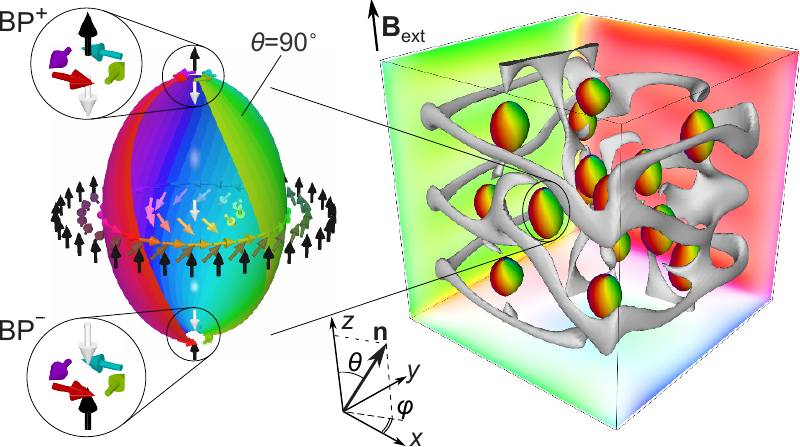}
    \caption{
    Left: Schematic representation of the spin texture of a magnetic globule. The isosurface indicates the position of the spins with $\theta=90^\circ$ while the color is defined according to the azimuthal $\varphi$ angle, $\mathbf{n}\!=\!(\sin\theta\cos\varphi, \sin\theta\sin\varphi, \cos\theta)$. 
    Insets show two Bloch points of opposite topological charge.
    Right: A cluster of globules in a chiral magnet shaped as a cube. 
    The confinement keeps the globules from moving apart.
    Isosurfaces of gray color denote spins with $\theta\!=\!5^\circ$ pointing nearly along $\mathbf{B}_\mathrm{ext}\!\parallel\! \mathbf{e}_z$. 
    }
\label{fig: globule visualisation}
\end{figure}

\begin{figure*}[ht]
    \includegraphics[width=0.9\linewidth]{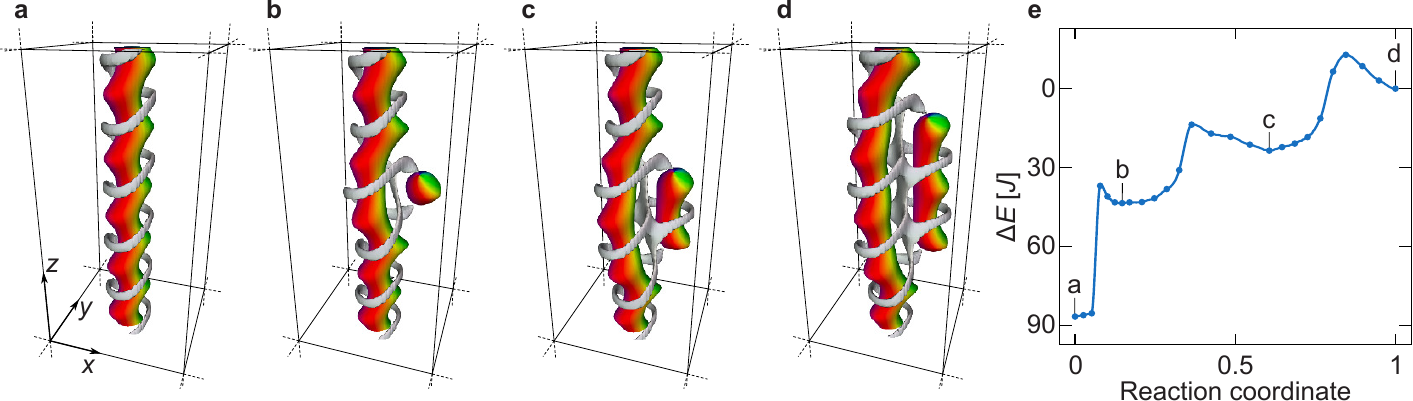}
    \caption{
    (a-d) 
    Stable 3D spin configurations representing inclusions in the conical phase in a bulk sample of an isotropic chiral magnet. 
    The inclusions are composed of a skyrmion tube and a magnetic globule of varying size coupled to it. 
    Isosurfaces of gray color denote spins with $\theta\!=\!5^\circ$ pointing nearly along $\mathbf{B}_\mathrm{ext}\!\parallel\! \mathbf{e}_z$. 
    (e) A minimum energy path between the elongated magnetic globule coupled to a skyrmion tube as shown in (d) and the isolated skyrmion tube shown in (a) calculated assuming periodic boundary conditions in all three directions and for $B_\mathrm{ext}\!=\!0.45B_\mathrm{D}$ with $L_\mathrm{D}=20a$.
    The states corresponding to the intermediate local minima are shown in (b) and (c). See also Supplementary Fig.~S2.1.
    }
\label{fig: bulk globule skyrmion}
\end{figure*}


The energy of the simulated system is described by an extended Heisenberg model for an isotropic chiral magnet defined by the following Hamiltonian~\cite{Han}: 
\begin{equation}
  E\!=\!- J \!\sum_{\left\langle ij\right\rangle }
   \mathbf{n}_i  \cdot  \mathbf{n}_j - 
  \!\sum_{\left\langle ij\right\rangle } 
  \!\mathbf{D}_{ij} \! \cdot \! [\mathbf{n}_i\! \times\!  \mathbf{n}_j]   - \! \mu_\mathrm{s} \mathbf{B}_\mathrm{ext}\!\sum_{i}\mathbf{n}_i,         
\label{E_tot}
\end{equation}
where $\mathbf{n}_i = \mathbf{\mu}_i/\mu_\mathrm{s}$ is the unit vector of the magnetic moment at lattice site $i$, the $\braket{ij}$ denote unique nearest-neighbour pairs, $J$ is the Heisenberg exchange constant and $\mathbf{D}_{ij}$ is the Dzyaloshinskii-Moriya vector defined as $\mathbf{D}_{ij}=D\mathbf{r}_{ij}$ with the scalar DMI constant $D$ and the unit vector $\mathbf{r}_{ij}$ pointing from site $i$ to site $j$, $\mathbf{B}_\mathrm{ext}$ is an external magnetic field. 
In order to keep the generality of the results presented below, as well as consistency with earlier studies, the size of the simulated domain is always given in terms of reduced units of distances with respect to $L_\mathrm{D}=2\pi a J/D$ -- the lowest period of a spin spiral in the continuous limit ($J\!\gg\!D$), where $a$ is the cubic lattice constant and reduced units of external magnetic field are given with respect to $B_\mathrm{D}= {D^2}/(\mu_\mathrm{s}J)$ -- the critical field at which the bulk system reaches the field polarized ferromagnetic state.
We do not take into account the influence of the demagnetization fields and thus describe the phenomena at the level of the basic (minimal) model.
It is known that this model~(\ref{E_tot}) describes well fundamental properties of chiral magnets and even has predictive power, while the addition of dipolar interactions can help achieve closer quantitative agreement with experiments, see for instance~\cite{Rybakov_15} and~\cite{Zheng_18}.

The magnetic texture of the configuration composed of two coupled singularities is illustrated in Fig.~\ref{fig: globule visualisation}.
Because the isosurface $n_z\!=\!0$ of the corresponding 3D spin texture has a nearly ellipsoidal shape with the singular points located on opposite poles, we refer to this object as magnetic globule (MG).
Because of an opposite topological charge of the BPs, the MG can, to a certain extent, be considered as a dipole.
Note that due to the presence of discontinuities, such an object can not be classified by the Hopf invariant~\cite{Bott_Tu}.

Despite the fact that the MG as a whole represents a topologically trivial object, a sufficiently large energy barrier is present under certain conditions to prevent its collapse via annihilation of the BPs.
For instance, as shown in Fig.~\ref{fig: globule visualisation}, a cluster of MGs can be stabilized due to confinement effects.
In this case, the interactions between the MGs makes such a configuration stable.
Below, we show that a single MG can also be stabilized in the interior of a crystal due to its coupling to textures representing defects or distortions in the ground state.

For a moderately strong external magnetic field, $B_\mathrm{ext}<B_\mathrm{D}$, the global energy minimum for~(\ref{E_tot}) corresponds to the conical state -- a helical spin spiral with $\mathbf{k}\parallel\mathbf{B}_\mathrm{ext}$ and magnetization tilted towards the direction of $\mathbf{B}_\mathrm{ext}$.
However, in real crystals the magnetic texture is not perfectly ordered but rather contains various defects, distortions or inclusions. 
As has been shown earlier, a skyrmion tube (SkT) represents one such stable inclusion within the conical phase~\cite{Rybakov_16,Leonov_16}.
Figure~\ref{fig: bulk globule skyrmion}a shows an isolated SkT within the conical phase, reflected by the spiraling distortion of the corresponding isosurfaces of $\mathbf{n}\parallel\mathbf{B}_\mathrm{ext}$ and $\mathbf{n}\perp\mathbf{B}_\mathrm{ext}$.
The states shown in Fig.~\ref{fig: bulk globule skyrmion}b-d represent stable configurations that include an MG of varying size coupled to the SkT.
The minimum energy path (MEP) between these configurations is shown in Fig.~\ref{fig: bulk globule skyrmion}e.
It was calculated using the geodesic nudged elastic band (GNEB) method~\cite{bessarab_method_2015} implemented in the \textit{Spirit} framework~\cite{spirit}.
The MEP clearly shows the presence of an energy barrier between the intermediate configurations, indicating that the MG attached to the SkT indeed represents a metastable state that cannot be destroyed by small excitations.
The distance $d_\mathrm{BP}$ between the poles of the elongated MGs is quantized with a pitch of $\sim L_\mathrm{D}$, while the lowest energy state always corresponds to the state with the smallest $d_\mathrm{BP}\lesssim  L_\mathrm{D}$ as in b, which may slightly vary with the value of applied field.
The energy of such a coupled state increases approximately linearly with $d_\mathrm{BP}$ and tends to the energy of two coupled infinitely long skyrmion tubes.
It has been shown recently that such coupled states appear due to the attractive interaction between skyrmion tubes at $B_\mathrm{ext}<B_\mathrm{D}$~\cite{Leonov_16,Loudon_2017,Du_18}.
A single SkT can host a number of MGs coupled to it as shown in Fig.~\ref{fig: bulk globule lattice}a and in Supplementary  Fig.~S2.2.

\begin{figure}[!b]
	\includegraphics[width=0.9\linewidth]{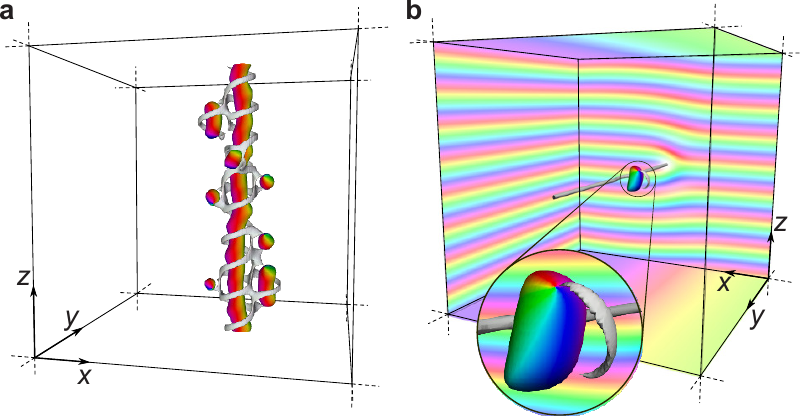}
    \caption{
(a) A cluster of stable globules in a bulk sample coupled to an isolated skyrmion tube embedded within a conical phase.
(b) Isolated globule stabilised at an edge dislocation within the bulk conical phase. 
Four surfaces of the simulated cube are shown to visualise the conical phase and the dislocation line. 
In (a) and (b) the isosurfaces show where the vector field is pointed in $xy$-plane (colored) and along positive direction of $z$-axis (gray), $\mathbf{B}_\mathrm{ext}\!\parallel\!\mathbf{e}_z$.
}
\label{fig: bulk globule lattice}
\end{figure}

\begin{figure*}[!bt]
\includegraphics[width=0.9\linewidth]{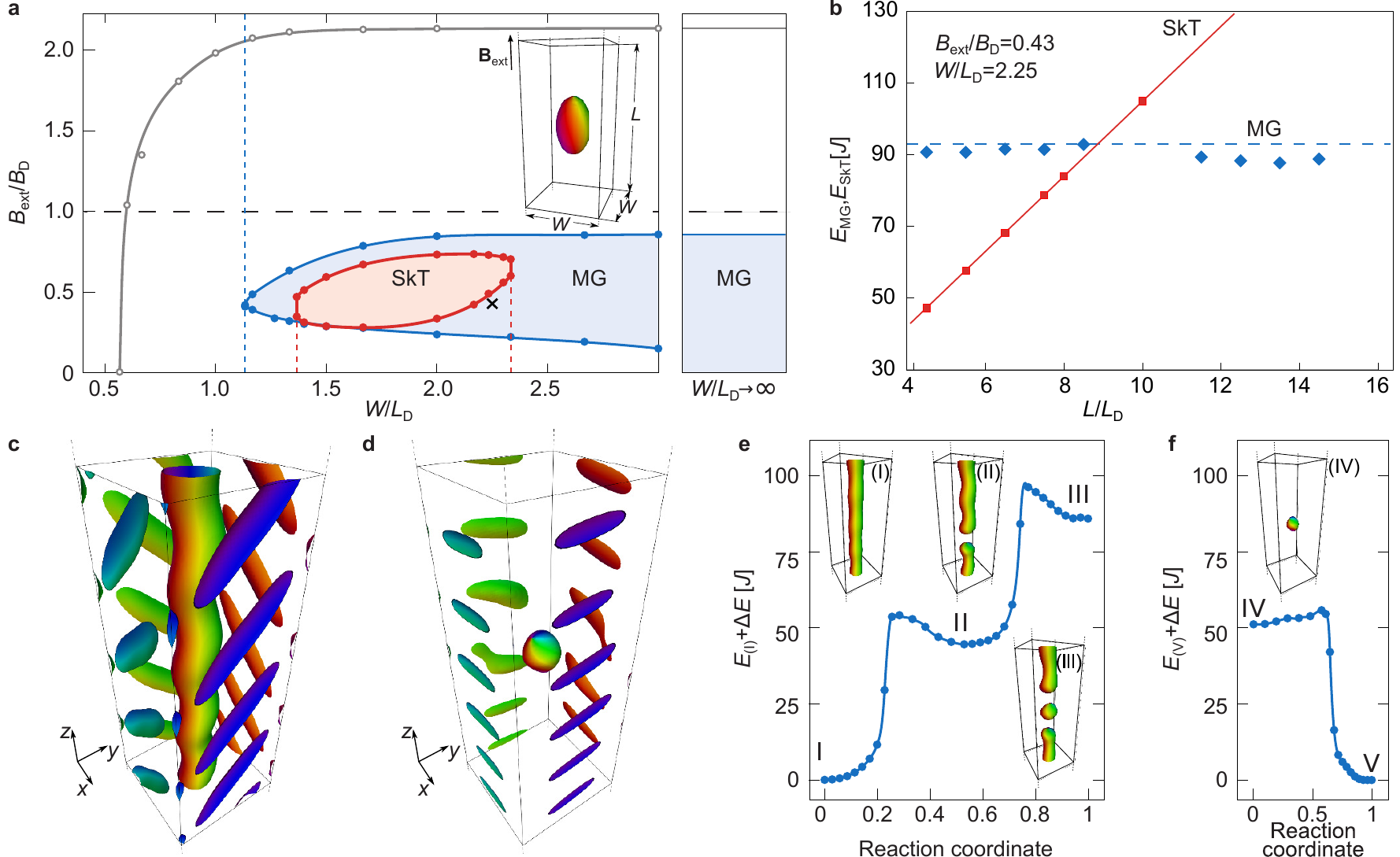}
\caption{
(a) Phase diagram of magnetic states in a nanowire of width $W$ and an external magnetic field $B_\mathrm{ext}$ applied along the nanowire. 
The red region corresponds to a single SkT ground state, while the stability range for the SkT is denoted by the solid gray line (open circles). 
The blue region corresponds to the stability of the isolated magnetic globule.  
Note, $W/L_\mathrm{D} \to\infty$ denotes convergence to the case of a semi-infinite crystal with a single surface representing a flat boundary between the crystal and vacuum.
$\times$ shows the parameter values chosen for (b-f).
(b) The self-energy of the MG $E_\mathrm{MG}$ and the SkT $E_\mathrm{SkT}$ with respect to the energy of the conical ground sate $E_\mathrm{cone}$ as function of the length $L$ of the nanowire.
The vertical dashed line represents the self-energy of the MG in an infinitely long nanowire, $E_\mathrm{MG}\approx93~J$, estimated using special boundary conditions as used for Fig.~\ref{fig: bulk globule skyrmion}b (see also \cite{supplement}).
(c) and (d) illustrate isosurfaces of the SkT and MG, showing a different periodicity in the background of conical phase with edge modulations.
(e,f) The minimum energy path, at $W=2.25L_\mathrm{D}$ and $B_\mathrm{ext}=0.43B_\mathrm{D}$, for the transition from an infinite skyrmion tube (I) to an isolated magnetic globule (IV) and then to the conical state with edge modulations (V).
More intermediate states are shown for $W=2L_\mathrm{D}$ and $B_\mathrm{ext}=0.45B_\mathrm{D}$ in Supplementary Fig.~S2.3.
Note that $L_\mathrm{D}=30a$ for (a-d), but $L_\mathrm{D}=20a$ for (e,f) due to computational constraints. As a result, there are some differences between (b) and (f).
}
\label{fig: globule nanowire}
\end{figure*}

Edge dislocations represent another type of inclusion that can appear within a conical phase. They belong to the family of native defects emerging in strip-like states in magnetism~\cite{dussaux2016} and beyond~\cite{harrison2000}.
We find that an edge dislocation can provide stability for the MG in the interior of the crystal as shown in Fig.~\ref{fig: bulk globule lattice}b.
In order to simulate such a single half plane defect in a  crystal, special boundary conditions are applied to the simulated domain ~\cite{supplement}. 
We find that the range in magnetic field strength where an MG coupled to an edge dislocation is stable includes at least from $0.31B_\mathrm{D}$ to $0.83B_\mathrm{D}$.

The results presented above clearly show that inclusions and inhomogeneities typically occurring in the conical phase, such as SkTs and edge dislocations, can provide stability to MGs in the interior of crystals. 
Therefore, one can expect that MGs play a significant role in the thermodynamic and transport properties of helimagnets.
However, in order to directly observe nanoscale textures in an experimental setup, such as X-ray or transmission electron microscopy (TEM), the sample needs to be small enough to be transparent to X-ray photons or high energy electrons.
Below we show that for samples of finite size, the stability of MGs is naturally provided by coupling with edge modulations inherent in chiral magnets~\cite{Meynell_14, Du_15, Jin_17}.

Figure~\ref{fig: globule nanowire}a shows a diagram of magnetic states calculated for a long nanowire in an external magnetic field applied along its main axis.
We find a surprisingly large range of applied field strength and nanowire width where a single MG remains stable.
When the applied field approaches $B_\mathrm{ext}\!=\!B_\mathrm{D}$ the conical phase is suppressed.
Similar to chiral bobbers, the saturation of the conical phase at high magnetic field limits the stability of the MG and leads to its collapse.
Similar to the case of an MG attached to an edge dislocation in a bulk crystal, the upper bound for the stability of the MG is $B_\mathrm{ext}\approx0.86B_\mathrm{D}$ and remains nearly constant for $W/L_\mathrm{D}>2$, but decreases as the wire is made thinner.
There is also a lower limiting field below which the MG disappears via the escape of the BPs from the system through the free edges.
The value of the lower bounding field decreases gradually with increasing width of the wire, $W$.
We estimate the critical width of the nanowire to be $W\approx10L_\mathrm{D}$, above which the MG may remain stable even at zero magnetic field. 

In the whole range where the MG is stable its energy is higher than that of the conical phase.
Within this range, there is a region where the energy density of a single SkT along the wire is lower than that of the conical phase, meaning that it represents the ground state of the system~\cite{Charilaou}.
Outside this region the SkT remains stable up to high $B_\mathrm{ext}$, see the gray curve in Fig.~\ref{fig: globule nanowire}a.
Note, in the spin lattice model the critical field of the skyrmion collapse is defined by the ratio of internal coupling parameters, $J/D$  while in the continuum limit the collapse does not occur even if $B_\mathrm{ext}\!\rightarrow\!\infty$~\cite{Doring}.
Fig.~\ref{fig: globule nanowire}b shows the dependence of the total energy of the SkT and MG as a function of the length $L$ of the wire with respect to the ground state which for the chosen parameters corresponds to the conical phase with edge modulations.
Since the SkT occupies the whole volume of the wire, its energy increases linearly with $L$.
On the other hand, the energy of the MG as a 3D-localized state remains approximately constant.
The presence of an MG introduces local distortion to the ground state.
In the calculations with periodic boundary conditions, it leads to the fact that for finite $L$ the energy of the MG may somewhat deviate from the self-energy of the MG when $L\rightarrow\infty$ (dashed line).
The distortions introduced by the SkT and MG in the nanowire are visualized in Fig.~\ref{fig: globule nanowire}c,d, where edge modulations have distinctly different periodicity.
Noticeably, such behavior of the self-energy of the SkT and MG applies to a broad range of parameter values.
In particular, for any $B_\mathrm{ext}$ and any $W/L_\mathrm{D}$ where the MG is stable, and the ground state is the conical phase with edge modulations (the blue region in Fig.~\ref{fig: globule nanowire}a), there is always a certain critical $L$ above which the MG is lower in energy than the SkT.
This can lead to a transition from SkT to MG.

To illustrate the possibility of such a transition, we calculated the MEP from a homogeneous SkT (I) to a single MG (IV) and then to the conical state with edge modulations (V).
In Fig~\ref{fig: globule nanowire}e and f we show only those parts of MEP which remain approximately invariant as the length of the nanowire is varied
The path between states (III) and (IV) depends on the length of the nanowire and is therefore not shown. There is a barrier on this stretch of path, and then the energy gradually drops while SkTs with one end, which are located above and below MG, move to the opposite ends of the wire (see also Supplementary Fig.~S2.3).
According to the above, the energy difference and thereby energy gain of the transition between state (I)
and (IV) depends on the length of the wire and for a sufficiently long wire, $L\!\gg\! L_\mathrm{D}$, and appropriate value of $B_\mathrm{ext}$ the condition $E_\mathrm{(I)}\!\gg\!E_\mathrm{(IV)}$ can be satisfied.

The energy barrier of $\sim 5J$ for the collapse of the MG is on the same order of magnitude as the energy barrier for the chiral bobber estimated with the same method in Ref.~\onlinecite{Rybakov_15}.
On the other hand, the height of the energy barrier is not the only factor determining the stability of localized states~\cite{bessarab_lifetime_2018}. 
The pre-exponential factor in the Arrhenius expression for the lifetime, reflecting the relative entropy of the 
transition state and the initial state also needs to be evaluated.
The estimation of the lifetime of the metastable states depicted in Fig.~\ref{fig: globule nanowire}e is a topic of future studies.

In conclusion, we have shown that in a wide class of isotropic chiral magnets a three-dimensionally localized state composed of a pair of Bloch points can be stable.
By calculating minimum energy paths, we have shown that a finite energy barrier protects such magnetic globules from collapsing.
In the interior of a crystal, an MG remains stable when it is located near a defect line of the conical phase or is coupled to a skyrmion tube.
We have shown that for a wide range in magnetic field strength and parameter values, in both in extended and  confined systems, magnetic globules can be stable as isolated objects.

In order to experimentally confirm the existence of MGs, the most promising avenue seems to be the partial or full reconstruction of the 3D magnetic configuration inside a crystal using some magnetic imaging technique, such as off-axis electron holography~\cite{Jin_17,Midgley_09}, the X-ray magnetic circular dichroism~\cite{Col_14} or X-ray vector nanotomography~\cite{Donnelly_17}.
According to recent studies, it is expected that stable magnetic textures containing singularities may also be confirmed with magnetoresistive measurements~\cite{Du_14,Du_15,redies_distinct_2018}.

\section*{Acknowledgments}
G.\,P.\,M.\ and H.\,J. acknowledge funding from the Icelandic Research Fund (grant no. 185405-051).
The work of F.\,N.\,R. was supported by the Swedish Research Council Grant No. 642-2013-7837 and by G\"{o}ran Gustafsson Foundation for Research in Natural Sciences and Medicine. 
The work of N.\,S.\,K. was supported by Deutsche Forschungsgemeinschaft (DFG) via SPP 2137 ``Skyrmionics'' Grant No. KI 2078/1-1.





\clearpage
\pagebreak
\onecolumngrid

\setcounter{equation}{0}
\setcounter{figure}{0}
\renewcommand{\thesection}{S\arabic{section}.}
\renewcommand{\thefigure}{S\arabic{section}.\arabic{figure}}
\renewcommand{\thetable}{S\arabic{section}.\arabic{table}}
\renewcommand{\theequation}{S\arabic{section}.\arabic{equation}}

{\centerline {\bf  \large Supplementary Material for ``Coupled quasimonopoles in chiral magnets''}}

\vskip 0.3 true cm

\centerline {Gideon P. M\"uller,$^{1,2,3}$ Filipp N. Rybakov$^{4}$}
\centerline { Hannes J\'onsson,$^{2}$ Stefan Bl\"ugel$^{1}$ and Nikolai S. Kiselev,$^{1}$}

\vskip 0.1 true cm
\centerline {\it $^1$Peter Gr\"unberg Institut and Institute for Advanced Simulation, Forschungszentrum J\"ulich and JARA, 52425 J\"ulich, Germany}
\centerline {\it $^2$Science Institute and Faculty of Physical Sciences, University of Iceland, 107 Reykjav\'{i}k, Iceland}
\centerline {\it $^3$Department of Physics, RWTH Aachen University, 52056 Aachen, Germany}
\centerline {\it $^4$Department of Physics, KTH-Royal Institute of Technology, SE-10691 Stockholm,  Sweden}
 
\vskip 0.5 true cm

To estimate the stability of magnetic globules presented in this work we performed spin dynamics relaxation based on Landau–Lifshitz–Gilbert equation of motion and calculation of minimum energy paths with the geodesic nudged elastic band (GNEB) method.
These calculations, as well as the visualisation of the isosurfaces, have been performed with the \textit{Spirit} framework \cite{spirit_supp}.
For very large systems we performed direct energy minimization with nonlinear conjugate gradients method implemented for NVIDIA CUDA architecture~\cite{Rybakov_15_supp}.
The latter allowed us to perform simulations for systems of large size $\sim(10L_\mathrm{D})^3$ with $L_\mathrm{D}\sim30a$, composed of $\sim 10^7$ lattice sites.

\stepcounter{section}
\section{Boundary conditions for the bulk system}
\begin{figure*}[ht]
    \includegraphics[width=0.9\linewidth]{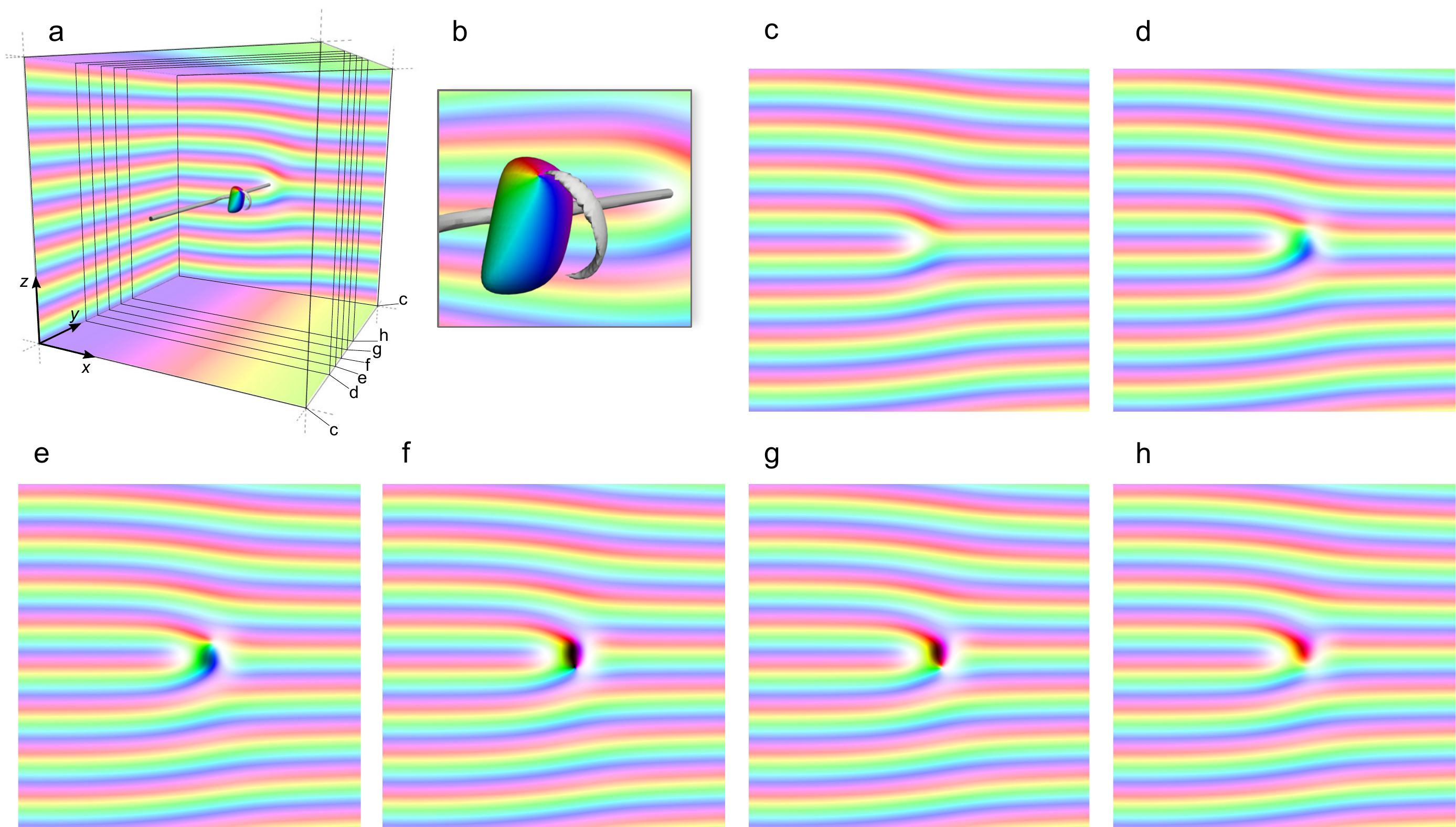}
    \caption{
    	Isolated magnetic globule in the the bulk crystal.
    	(a) 3D visualization of magnetic globule coupled to an edge dislocation in the conical phase (see Fig.~3 of the main text).
        (a,b) isosurfaces for $n_z\approx0$ and $n_z\approx1$
        (c)-(h) cross sections of the domain in the $zx$-plane near the center of the magnetic globule, as indicated in (a).
        }
\label{fig: supplement bulk globule dislocation}
\end{figure*}

Ordinary periodical or open boundary conditions cannot be applied in the case presented in Fig.~\ref{fig: bulk globule lattice} of the main text due to the broken symmetry of the spin texture.
This can be clearly seen in Fig.~\ref{fig: supplement bulk globule dislocation}.
Both the $x$- and $z$-direction cannot be treated with periodic boundary conditions.

In order to calculate physically correct states of a bulk crystal, but with non-periodical boundary conditions, a special scheme was applied.
In $y$-direction an ordinary periodical boundary conditions are applied.
In the $x$-direction, the spins at the boundary layers are pinned according to the analytical solution for the conical state at the corresponding applied magnetic field, $\theta=\arccos(B_\mathrm{ext}/B_\mathrm{D})$, $\varphi=2\pi z/L_\mathrm{D} + \varphi_0$.
In order to enforce the stability of the dislocation within the simulated domain, the values of $\varphi_0$ on the left and right boundary of the domain ($x_\mathrm{min}$ and $x_\mathrm{max}$) are set to be different with $\Delta\varphi_0=\pi$.   
The infinite continuation in $z$-direction is achieved by the following method. 
In ordinary periodical boundary conditions the upward neighbors for spins in the endmost upper layer, $z=z_\mathrm{max}$ are the spins in the endmost lower layer with $z=z_\mathrm{min}$.
Contrary to this, in our implementation, the spins in the layer with $z=z_\mathrm{max}-L_\mathrm{D}$ play the role of upward neighbors for the endmost upper layer.
Correspondingly for the endmost lower layer, $z=z_\mathrm{min}$ the downward neighboring spins are in the layer with $z=z_\mathrm{min}+L_\mathrm{D}$.
This enables the calculation of a single defect line in the strip-like phase, as shown in Fig.~3 in the main text.
From the above it is clear that the size of the simulated domain along the $z$-axis, $L_z$, has to be a multiple of $L_\mathrm{D}$. Moreover, $L_z$ should be large enough to host at least few periods of the spin spiral and material parameters have to be chosen such that $L_\mathrm{D}=N\cdot a$, where $N$ is an integer and $a$ is the lattice parameter.
In our simulations we used $L_z=8L_\mathrm{D}$ and $L_\mathrm{D}=32a$, meaning $256$ lattice sites along the $z$-axis.
For consistency, we used the same number of lattice sites in the other directions, $L_x=L_y=L_z$.



\setcounter{equation}{0}
\setcounter{figure}{0}

\stepcounter{section}
\section{Supplementary figures}

\begin{figure*}[ht]

\includegraphics[width=0.7\linewidth]{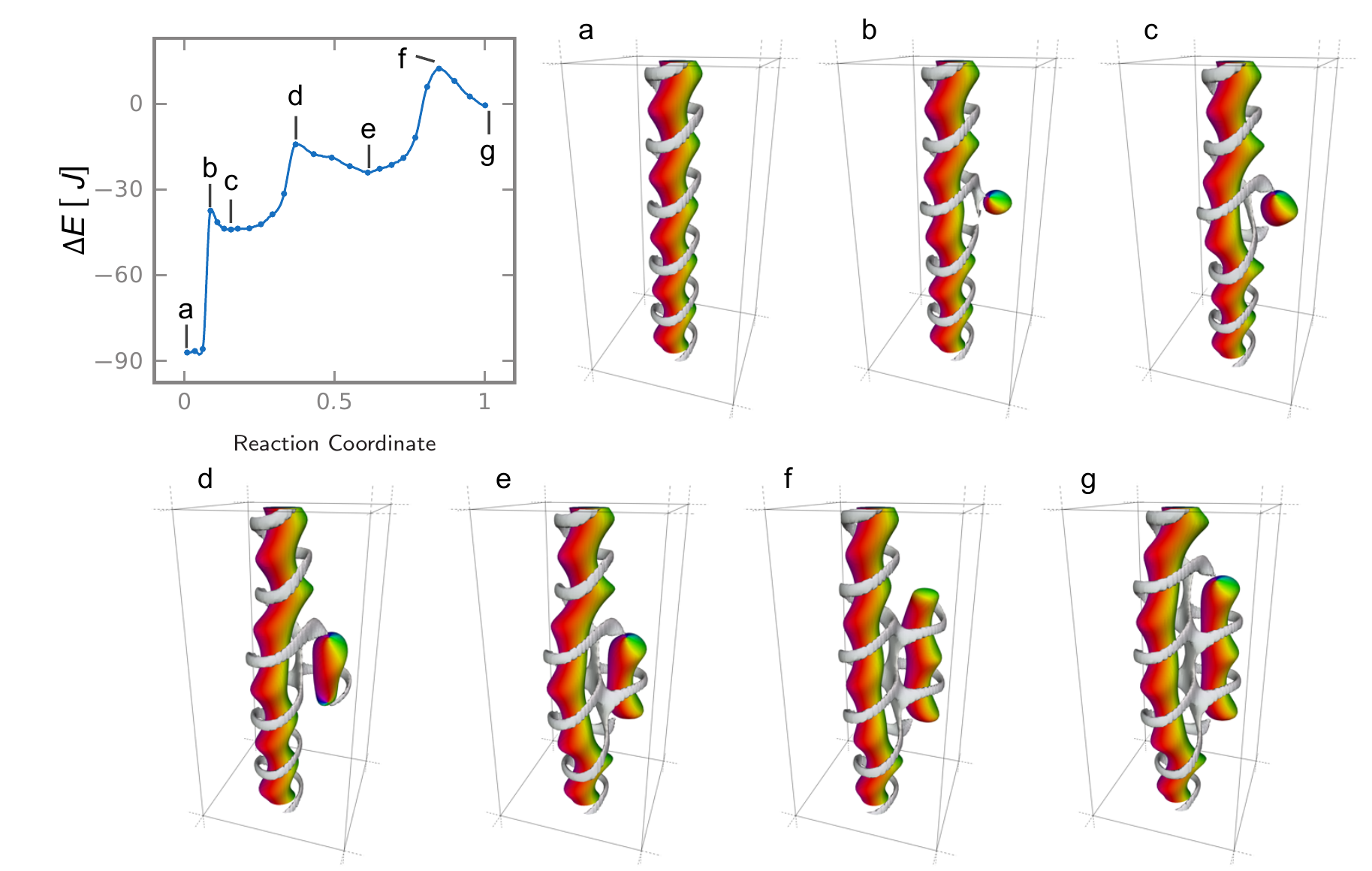}
\caption{
Minimum energy path and corresponding spin configurations of a bulk crystal containing an infinite skyrmion tube (a), showing the nucleation of a magnetic globule coupled to the skyrmion tube (c) and the subsequent elongation of the globule (e,g), shown in  in Fig.~1 of the main text.
The energy difference $\Delta E$ is with respect to the final coupled state (g).
The path was calculated in a system of sufficient size that the interaction via the periodical boundaries should not significantly influence the barriers.
The full set of stationary points of the minimum energy path is shown.
$B/B_\mathrm{D}=0.45$, $L_\mathrm{D}=20a$, $W=2L_\mathrm{D}$, $L_z=6L_\mathrm{D}$
}
    
\end{figure*}

\begin{figure*}[ht]  
    \begin{minipage}[b]{0.25\textwidth}
    \includegraphics[width=\linewidth]{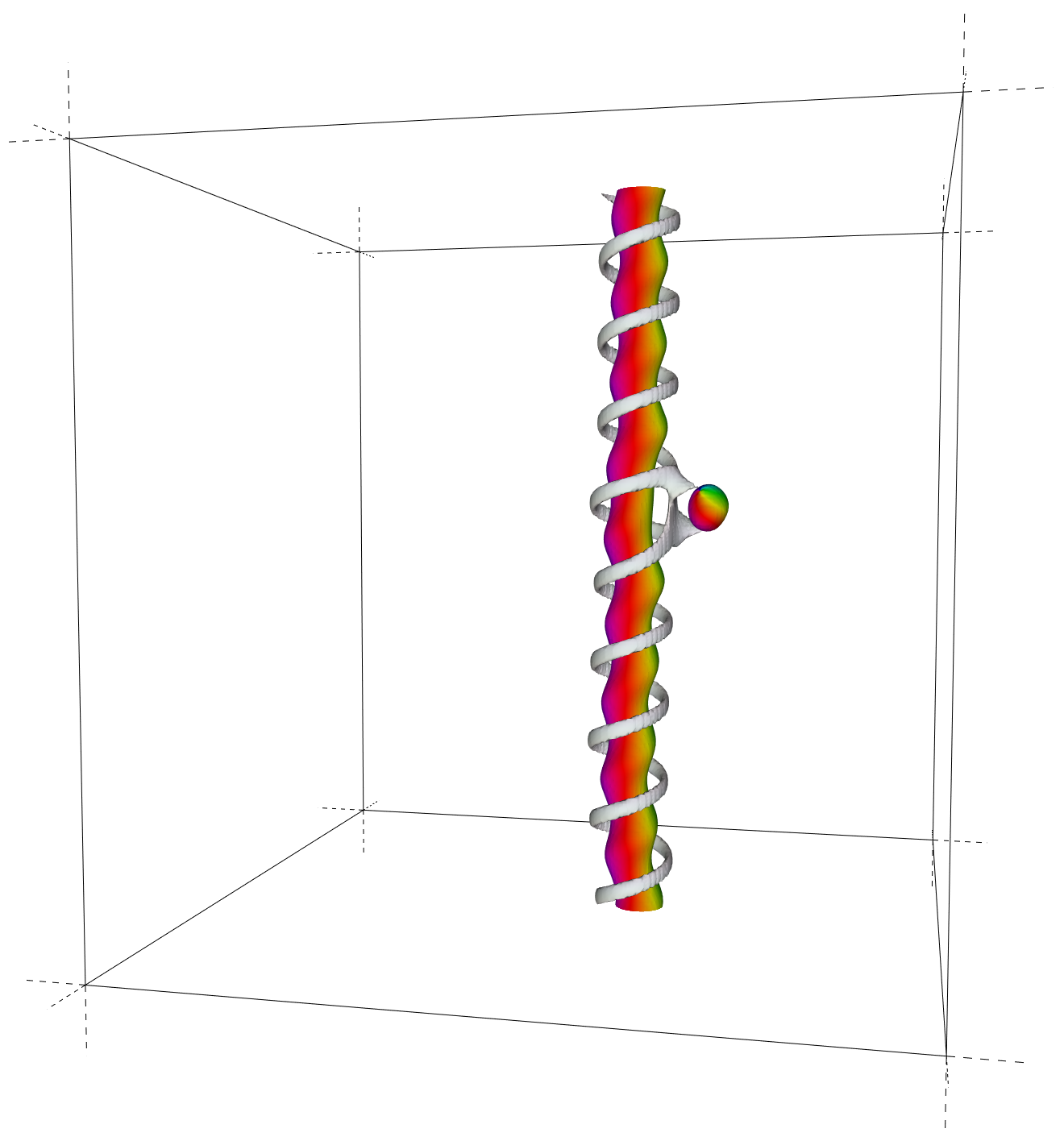}
    \end{minipage}
    \begin{minipage}[b]{0.25\textwidth}
    \includegraphics[width=\linewidth]{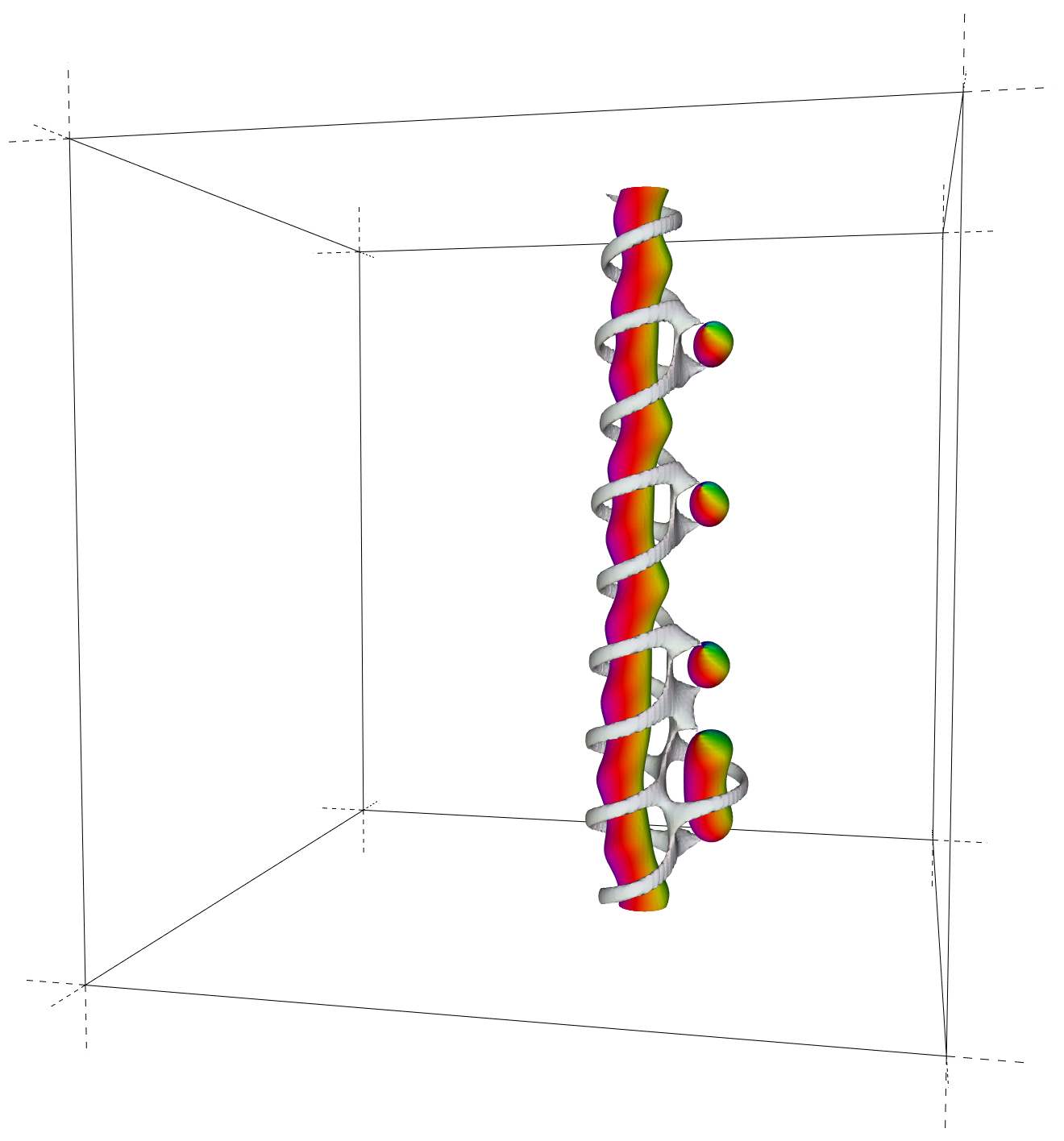}
    \end{minipage}
    \begin{minipage}[b]{0.25\textwidth}
    \includegraphics[width=\linewidth]{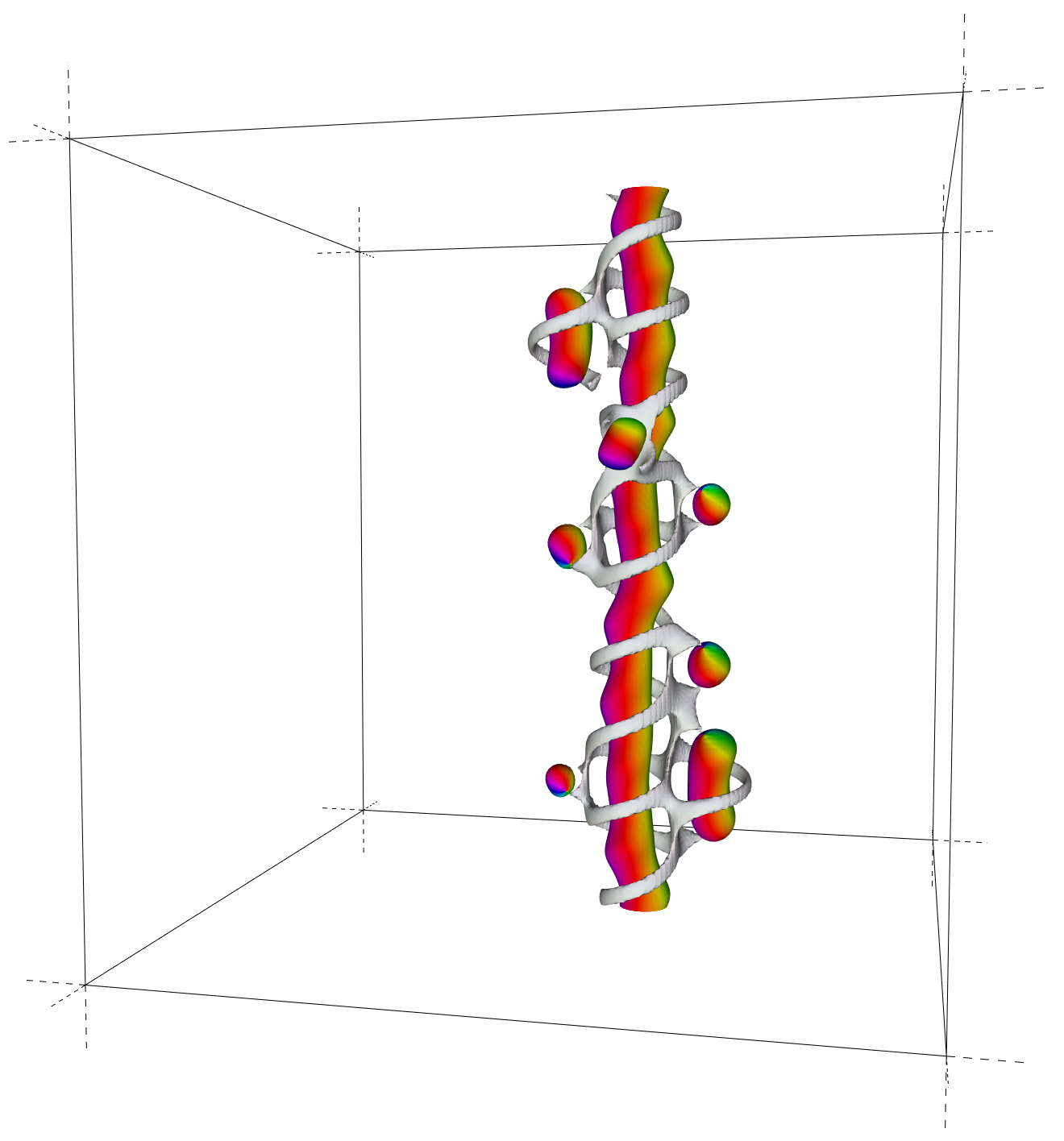}
    \end{minipage}
    \caption{
    	Globules coupled to a skyrmion tube in a cube of side length $9~L_\mathrm{D}$.
        A single globule may be stable due to it's coupling to the skyrmion tube, even inside an otherwise conical state without further inhomogeneities.
        It is shown that multiple globules, including elongated globules may be coupled to the same skyrmion tube.}
    
\end{figure*}

\begin{figure*}[ht]
    
    \includegraphics[width=0.7\linewidth]{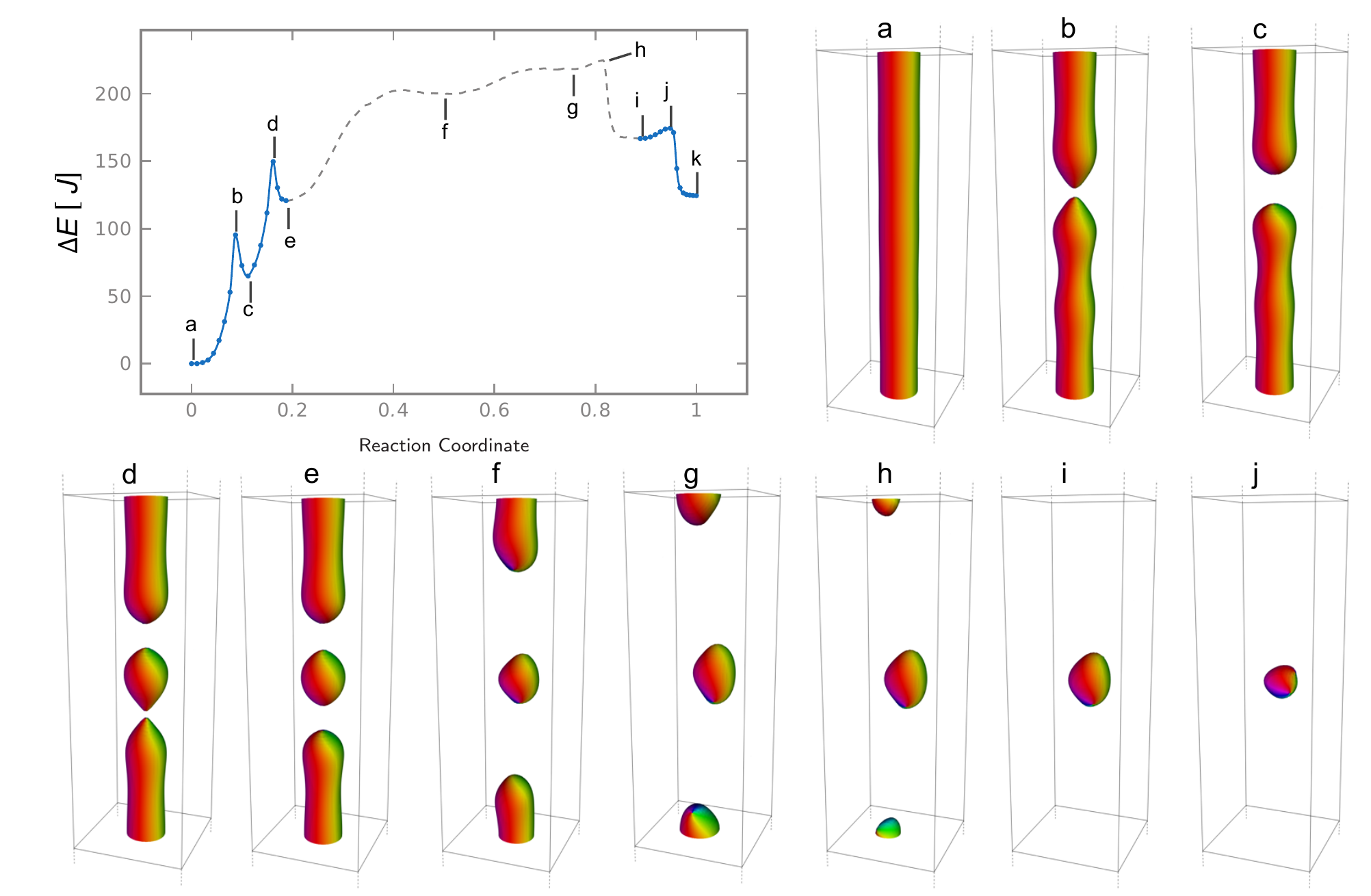}
    \caption{
    	Full set of the stationary points of the minimum energy path, analogous to Fig.~3 of the main text, at $W=2L_\mathrm{D}$ and $B_\mathrm{ext}=0.45B_\mathrm{D}$.
    	A nanowire is shown with the isosurfaces for $n_\mathrm{z}\!=\!0$.
    	Minimum energy path for the nucleation of a globule inside a skyrmion tube and the collapse of a globule.
        In order of appearance: infinite skyrmion tube, saddle point of first cut, metastable finite skyrmion tube, saddle point of second cut, globule coupled inside of skyrmion, normal and elongated globule together, two normal globules, saddle point for collapse of first globule, single globule, saddle point for collapse of the single globule.
        The wire is periodically continued in z-direction, meaning that the two skyrmion halves in (g) represent a second globule. The intermediate minimum appears due to the interaction between the Bloch points inside the wire in combination with their interaction over the periodical boundaries. Hence, we call (e-i) an artificial transition, which we show here only to illustrate the possible interactions between these singularities.
        }
    
\end{figure*}

\FloatBarrier

\end{document}